\documentclass[aps,showpacs,twocolumn]{revtex4}
\usepackage[pctex32]{graphics}
\begin{document}
\newcommand{\bstfile}{aps} 
\title{A New and Compact Sum-Over-States Expression without Dipolar Terms for Calculating Nonlinear Susceptibilities}
\author{Mark G. Kuzyk}
\address{Department of Physics and Astronomy, Washington State University \\ Pullman,
Washington  99164-2814}
\date{\today}

\begin{abstract}
Using sum rules, the dipolar terms can be eliminated from the commonly-used sum-over-states (SOS) expression for nonlinear susceptibilities.  This new dipole-free expression is more compact, converges to the same results as the common SOS equation, and is more appropriate for analyzing certain systems such as octupolar molecules.  The dipole-free theory can be used as a tool for analyzing the uncertainties in quantum calculations of susceptibilities, can be applied to a broader set of quantum systems in the three-level model where the standard SOS expression fails, and more naturally leads to fundamental limits of the nonlinear susceptibilities.
\end{abstract}


\maketitle


\vspace{1em}

\section{Introduction}

The sum-over-states (SOS) expression for the nonlinear-optical susceptibilities,\cite{orr71.01} which are expressed in terms of the matrix elements of the dipole operator, $-e x_{nm}$, and the energy eigenvalues, $E_n$, have been used extensively over the last 4 decades as a theoretical model of the nonlinear response as well as a tool for analyzing experimental dispersion data.  Indeed, the two-level model has guided the development of molecules with large second-order nonlinear-optical susceptibilities (also called hyperpolarizabilities).  Similarly, the SOS three-level model for the third-order susceptibility (also called the second hyperpolarizability) has led to an understanding of the nature of the states, and interactions between them, that yield the largest response.

While the SOS expression has reigned supreme for 4 decades, there are several critical issues that have never been addressed.  Because the expression is over-specified, redundant information is required to calculate nonlinear susceptibilities.  This redundancy not only leads to inefficiencies in the computational process; but, the SOS expression can lead to reasonable-looking results even when unphysical parameters are used as the input.  As such, the underlying physics of the nonlinear-optical response may be misinterpreted, leading to erroneous conclusions.

The sum rules are quantum mechanical identities that relate the dipole matrix elements and energies to each other; so, the SOS hyperpolarizability can be expressed in terms of a subset of the dipole matrix.\cite{kuzyk00.01}  In this work we show that all the dipolar terms can be eliminated to yield a simplified expression that is equivalent to the full SOS expression.  This theoretical result can be used as a tool for studying the nonlinear-optical response.  For example, differences between the full SOS expression and the dipole-free expression can be used to estimate the uncertainties in quantum calculations since such discrepancies are a sign that the sum rules have been violated (due to incorrect dipole matrix elements and energies, or truncation errors).  The approximate solution of a particle in a tilted box is analyzed in this way as an illustration.  More importantly, the three-level model of the dipole-free theory may be applicable when the standard three-level SOS expression fails, thus providing a theoretical tool that covers a broader base of quantum systems.  To test our new theoretical expression we show that for a clipped harmonic oscillator, for which analytical solutions to the Schr\"{o}dinger Equation can be computed, the two theories converge to the same results in the limit of an infinite-level model.  Finally, we shall show that the dipole-free expression more naturally leads to fundamental limits of the nonlinear susceptibilities.

\section{Theory}

In this section, we show how the SOS expression of the first and second hyperpolarizabilities are simplified using the sum rules.  The commonly-used SOS expression for any diagonal component of $\beta$ is given by:\cite{orr71.01}
\begin{eqnarray}
& & \beta_{xxx} (- \omega_{\sigma}; \omega_1, \omega_2) = -  e^3 \frac {1} {2} P(\omega_{\alpha}, \omega_{\beta}) \\ \nonumber
& &  \left[ {\sum_{n}^{\infty}} ' \frac {\left| x_{0n} \right|^2 \Delta x_{n0}} {E_{n0}(\omega_{\alpha}) E_{n0}(\omega_{\beta})} \right. \\ \nonumber
& + & \left. {\sum_{n }^{\infty}} ' {\sum_{m \neq n }^{\infty}} ' \frac {x_{0n} x_{nm} x_{m0}} {E_{n0} (\omega_{\alpha}) E_{m0} (\omega_{\beta}) }\right],
\label{beta}
\end{eqnarray}
where $e$ is the magnitude of the electron charge, $x_{nm}$ the $n,m$ matrix element of the position operator, $\Delta x_{n0} = x_{nn} - x_{00}$ is the difference in the expectation value of the electron position between state $n$ and the ground state $0$, $E_{n0} = E_n - E_0$ is the energy difference between the excited state $n$ and the ground state, $E_{m0} (\omega_{\beta}) \equiv E_{m0} - \hbar \omega_{\beta} $, and $\hbar \omega_{\beta}$ is the energy of one of the photons.  The primes indicate that the ground state is excluded from the sum and the permutation operator $P(\omega_{\alpha}, \omega_{\beta})$ directs us to sum over all six frequency permutations given by the Feynman Diagrams.  Since the dipole moment of the molecule is proportional to the position ($p_x = -e x$), we loosely call $x_{nm}$ the transition moment and $x_{nn}$ the excited state dipole moment.   The first term in Equation \ref{beta} is called the dipole term and the second term the octupolar term; and, as we shall see below, the dipole term can be expressed in terms of the octupolar one using the sum rules.  

The generalized Thomas-Kuhn sum rules are a direct consequence of the Schr\"{o}dinger Equation (without any approximations) and relate the matrix elements and energies to each other according to:\cite{kuzyk01.01}
\begin{equation}
\sum_{n=0}^{\infty} \left( E_n - \frac {1} {2} \left( E_m + E_p \right) \right) x_{mn} x_{np} = \frac {\hbar^2 N} {2m} \delta_{m,p},
\label{sumrule}
\end{equation}
where $m$ is the mass of the electron, and $N$ the number of electrons in the molecule.  The sum, indexed by $n$, is over all states of the system.  Equation \ref{sumrule} represents an infinite number of equations, one for each value of $m$ and $p$.  As such, we refer to a particular equation using the notation $(m,p)$.

To eliminate the dipole term, we consider the Equation $(m,p)$ with $m \neq p$:
\begin{equation}
\sum_{n=0}^{\infty} \left( E_{nm} + E_{np} \right) x_{mn} x_{np} =0 .
\label{sumrule m<>p}
\end{equation}
Equation \ref{sumrule m<>p} can be rewritten by explicitly expressing the $n=m$ and $n=p$ terms:
\begin{eqnarray}
& & \sum_{n=0(\neq p, \neq m)}^{\infty} \left( E_{nm} + E_{np} \right) x_{mn} x_{np} \\ \nonumber
& + & E_{mp}x_{mm}x_{mp} + E_{pm}x_{mp}x_{pp} = 0 .
\label{sumrule m<>p m,p removed}
\end{eqnarray}
Using $E_{mp} = -E_{pm}$ and the definition $\Delta x_{pm} = x_{pp}-x_{mm}$, Equation \ref{sumrule m<>p m,p removed} becomes,
\begin{equation}
\sum_{n=0(\neq p, \neq m)}^{\infty} \left( E_{nm} + E_{np} \right) x_{mn} x_{np} + E_{mp}x_{mp} \Delta x_{mp} = 0.
\label{sumrule m<>p m,p removed final}
\end{equation}

Setting $p=0$ in Equation \ref{sumrule m<>p m,p removed final} and solving for $\Delta x_{n0} \left| x_{0n} \right|^2$ after multiplying through by $x_{0m}$, we get
\begin{equation}
\Delta x_{m0} \left| x_{0m} \right|^2 = - {\sum_{n \neq m}^{\infty}} ' \frac {E_{nm} + E_{n0}} {E_{m0}}  x_{0m} x_{mn} x_{n0} .
\label{sumrule m<>p m,p removed dipole}
\end{equation}
Substituting Equation \ref{sumrule m<>p m,p removed dipole} with $m \leftrightarrow n$ into Equation \ref{beta}, we get the final result,
\begin{eqnarray}
& & \beta_{xxx} (- \omega_{\sigma}; \omega_1, \omega_2) = - \frac { e^3} {2} P(\omega_{\alpha}, \omega_{\beta}) {\sum_{n}^{\infty}} ' {\sum_{m  \neq n}^{\infty}} ' \\ \nonumber
& & \frac {x_{0n} x_{nm} x_{m0}} {E_{n0}(\omega_{\alpha}) E_{m0}(\omega_{\beta})} \left[ 1 - \frac {E_{m0}(\omega_{\beta}) \left(2E_{m0} - E_{n0} \right) } {E_{n0} E_{n0}(\omega_{\beta})}\right]. 
\label{beta contracted}
\end{eqnarray}
We call this form of $\beta$ the dipole-free expression or the reduced hyperpolarizability.  The second term in brackets is the dispersion term that results when the dipolar terms are eliminated.  In the standard SOS expression, the simplest approximation is the two-level model, with parameters $x_{10}$, $\Delta x_{10}$, and $E_{10}$.  The simplest approximation to Equation \ref{beta contracted} is the three-level model with parameters $x_{10}$, $x_{20}$, $x_{12}$, $E_{10}$, and $E_{20}$.  This is in contrast to the standard SOS expression, where the three-level model has two additional dipole terms.

It is important to note that while the dipole-free expression may seem to be less general than the common SOS one, when all states are included, it is fully equivalent.  Because the sum rules are a direct consequence of the Schr\"{o}dinger Equation, they can not be violated in any system, be it an atom, molecule, or crystal.  The SOS expression, in both forms, can be evaluated for unphysical values of the matrix elements - yielding nonsensical values of the hyperpolarizability.  However, in its nonrestricted form, there is more room for introducing errors.  The restriction imposed on the SOS expression used to get the dipole-free equation prevents certain unphysical combinations of dipole and octupolar terms, so in non-truncated form, is more robust.  For example, the lowest truncated-state model in the standard SOS expression is the two-level model, which only describes transitions in which the dipole moment changes between these two states.   However, it ignores all octupolar terms.  On the other hand, the dipole-free expression - when truncated - can approximate molecules with octupolar character \cite{Joffr92.01,fiori95.01,diazg94.03} as measured with hyper-Rayleigh scattering;\cite{clays91.01,olbre20.01} as well as dipolar terms, which are implicitly taken into account by the extra dispersion term in the reduced hyperpolarizability, as given by Equation \ref{beta contracted}.

All higher-order nonlinear susceptibilities can be treated in the same way.  As an illustration, we briefly consider the third-order susceptibility.  For any diagonal component of $\gamma$, the second hyperpolarizability, along the $x$-direction is given by:
\begin{equation}
\gamma_{xxxx} =  {\sum_{n,m,l}^{\infty}}' \frac {x_{0n} \bar{x}_{nm} \bar{x}_{ml} x_{l0}} {D_1(n,l,m)} - {\sum_{n,m}^{\infty}}' \frac {x_{0n} x_{n0} x_{0m} x_{m0}} {D_2(n,m)} ,
\label{gamma}
\end{equation}
where $D_i(n,m,...)$ are energy denominators and $n,m,...$ are arguments that show which energies are represented (i.e. $E_{n0}$, $E_{m0}$, ...). In analogy to $\beta$ as given by Equation \ref{beta}, the denominators are of the form $D_1(n,l,m) = \hbar^3 E_{n0}(\omega_{\alpha}) E_{m0}(\omega_{\beta}) E_{l0}(\omega_{\delta})/4e^4$ and are most easily determined using Feynman Diagrams for the particular phenomena of interest.  There are two terms in Equation \ref{gamma} that depend on the dipole moment, which can be expressed as,
\begin{equation}
T_1 = {\sum_{n,m \neq n }^{\infty}}' \frac {x_{0n} x_{nm} \Delta x_{m0} x_{m0}} {D_1(n,m,m)} ,
\label{gamma1}
\end{equation}
and
\begin{equation}
T_2 =  {\sum_{n}^{\infty}}' \frac { \left| x_{0n} \right|^2 \Delta x_{n0}^2} {D_1(n,n,n)} .
\label{gamma2}
\end{equation}

Using Equation \ref{sumrule m<>p m,p removed final} with $p=0$ and $n=l$, Equation \ref{gamma1} becomes
\begin{equation}
T_1 = - {\sum_{n,m \neq n ,l \neq m}^{\infty}}' \hspace{1em} \frac {x_{0n} x_{nm} x_{ml} x_{l0}} {D_1(n,m,m)} \cdot \left( \frac {E_{l0} + E_{lm}} {E_{m0}} \right) .
\label{gamma1b}
\end{equation}
Similarly, Equation \ref{gamma2} can be written as
\begin{eqnarray}
T_2 & = & {\sum_{n,m \neq n,l\neq m}^{\infty}}' \hspace{1em} \frac {x_{0n} x_{nm} x_{ml} x_{l0}} {D_1(n,n,n)} \\ \nonumber 
& \times & \left( \frac {E_{l0} + E_{lm}} {E_{m0}} \right)  \cdot  \left( \frac {E_{n0} + E_{nm}} {E_{m0}} \right) .
\label{gamma2b}
\end{eqnarray}
Using Equations \ref{gamma1b} and \ref{gamma2b},  Equation \ref{gamma} for $\gamma$ can be written in dipole-free form in analogy to Equation  \ref{beta contracted}.  Given the algebraic messiness of the result, it is not presented here.

\section{Applications}

In this section, we apply the dipole-free theory to several problems to show both its usefulness; and, to confirm that in the limit of including all states in the SOS expression, our new theory and the standard SOS expression converge.  Furthermore, we show that when approximate wavefunctions are used, such as the particle in a tilted box, the two results do not converge, showing how such a comparison can be used to 
assess the accuracy of the calculated nonlinearities.  In addition, we show how the reduced hyperpolarizability leads to a more elegant calculation of the fundamental limits of nonlinear susceptibilities.

First, we apply the theory to the calculation of the fundamental limits of the hyperpolarizability.\cite{kuzyk00.02,kuzyk01.01,kuzyk00.01,kuzyk03.01,kuzyk03.02,kuzyk04.02}  The issue of fundamental limits has been an important one since it guides the applied researcher in making better materials and devices while giving the theorist a method of understanding the nuances of what makes a large nonlinear-optical response.\cite{Champ05.01,kuzyk05.01} Using the new theory, we show that the results are the same; but, leads to a more elegant approach that illustrates the equivalence of viewing a molecule in terms of the standard expression that includes dipole terms or the reduced form in terms of octupolar terms.  {\em This provides an important paradigm shift} in the sense that the sum rules show that the two limiting cases are closely related while the general nonlinear-optics community operates on the assumption that the two are independent.

Second, the clipped harmonic SOS expression - for which exact analytical wavefunctions can be calculated - the SOS expression and the dipole-free theory presented here are compared.  The fact that the two converge shows that the two expressions are identical in the infinite-level model.  It is common for theorists and experimentalists to use limited-state models.  The two- and three-level models have been successfully applied to understanding the dispersion and magnitude of the second and third-order susceptibilities, but clearly, such a simplified view can not be universally correct.  Indeed, the three level model for the SOS expression and the dipole free one are totally different functions with different dispersion.

There are two important ramifications of this observation.  First, in cases where the standard truncated SOS expression is inconstant with observation, the dipole-free expression may be more appropriate.  As such, the dipole-free expression provides researchers with a tool to study a class of systems that was previously unaccessible.  Octupolar molecules may be one such class.  Secondly, a comparison of the dipole-free expression with the standard SOS theory can be used to estimate whether or not a theoretical calculation has converged without the need for including more and more numbers of states to test for convergence.  So, the theory presented here can save on computational time while providing the theorist with another tool.

Third, $\beta$ of a particle in a tilted box is analyzed.  Since this system is solved with perturbation theory, it is possible to study how small errors in the matrix elements affect the nonlinear response predicted by the two theories.  Most quantum chemical calculations yield only approximate wavefunctions, so this example shows how the two expressions can yield different results even in an infinite state model.  It must be stressed that there are no means for determining which calculation yields the ``correct" $\beta$ values that would be found experimentally.  Perhaps more importantly, a comparison between the two theories provides an estimate of the uncertainty of the calculation: The calculated values of the nonlinearity can only be trusted to within the range between the two values.  If both converge to greatly differing values in the infinite-state limit, then this suggests that the wavefunctions may be unphysical.

\subsection{Fundamental Limits}

To illustrate the usefulness of the dipole-free SOS expression, we apply it to calculate the fundamental upper limit of $\beta$\cite{kuzyk00.02,kuzyk01.01,kuzyk00.01,kuzyk03.01,kuzyk03.02,kuzyk04.02}.  We start with the sum rules $(0,0)$ and $(1,1)$ truncated to three levels, which yield
\begin{equation}
\left| x_{02} \right| = \sqrt{E \left( \left| x_{01}^{MAX} \right|^2 - \left| x_{01} \right|^2 \right) },\label{x02}
\end{equation}
and
\begin{equation}
\left| x_{12} \right| = \sqrt{\frac {E} {1-E} \left( \left| x_{01}^{MAX} \right|^2 + \left| x_{01} \right|^2 \right) },\label{x12}
\end{equation}
respectively, where
\begin{equation}
\left| x_{10}^{MAX} \right|^2 = \frac {\hbar^2} {2mE_{10} } N .
\label{groundsumrule}
\end{equation}
Substituting Equations \ref{x02} and \ref{x12} into Equation \ref{beta contracted} in the off-resonance limit ($\omega_{\alpha} = \omega_{\beta} = 0$), we get 
\begin{equation}
\beta = 6 \sqrt{\frac {2} {3}} e^3 \frac {\left| x_{10}^{MAX} \right|^3} {E_{10}^2} G(X) f(E) = \beta_0 G(X) f(E) ,\label{betafG}
\end{equation}
where
\begin{equation}
f(E) = (1-E)^{3/2} \left( E^2 + \frac {3} {2} E + 1 \right),\label{DEFf(E)}
\end{equation}
and
\begin{equation}
G(X) = \sqrt[4]{3} X \sqrt{\frac {3} {2} \left( 1 - X^4\right)},\label{defG(X)}
\end{equation}
where $X = x_{10} / x_{10}^{MAX}$ and $E = E_{10} / E_{20}$.

$G$ and $X$ are maximum at $G(\sqrt[-4]{3}) = 1$ and $f(0) = 1$, yielding,
\begin{equation}
\beta_{MAX} = \beta_0 f(0) G(\sqrt[-4]{3}) =  \sqrt[4]{3} \left( \frac {e \hbar} {\sqrt{m}} \right)^3 \left[ \frac {N^{3/2}} {E_{10}^{7/2}} \right] . \label{betaMAX3L}
\end{equation}
This is identical to the results from the usual sum-over-states expression; however, the calculation is much more concise and elegant because the dipolar term does not need to be considered.

\subsection{The Clipped Harmonic Oscillator}

In this section, we test the dipole-free SOS expression by comparing the results it gives with the standard SOS expression for a potential in which the Schr\"{o}dinger Equation can be solved analytically.  This approach ensures that the energies and dipole matrix elements are physically sound and that pathologies or inaccuracies normally inherent in approximation techniques are avoided.  We use the exact solution to the clipped harmonic oscillator (CHO) (where the potential is harmonic for $x>0$ and infinite for $x<0$) since it is the simplest case of an asymmetric potential that yields a large hyperpolarizability that is in fact near the fundamental limit.\cite{Tripa04.01} The matrix elements of the position operator of the clipped harmonic oscillator (CHO) are given by,
\begin{equation}
x_{mn} = x_{10}^{MAX} g_{mn},
\label{CHO-xmn}
\end{equation}
where the dimensionless matrix $g_{mn}$ is defined by
\begin{equation}
g_{mn} = \frac {2} {\sqrt{\pi}} (-1)^{((m+n)/2)} \cdot \left( \frac {2} {(m-n)^2 - 1 } \right) \cdot \left( \frac {m!! n!!} {\sqrt{m!n!}} \right) ,
\label{CHO-gmn}
\end{equation}
where $m$ and $n$ are restricted to the odd integers.  The energy of state $n$ is given by
\begin{equation}
E_n = \hbar \omega_0 \left( n + \frac {1} {2} \right) .
\label{CHO-energy}
\end{equation}

\begin{figure}
\scalebox{1}{\includegraphics{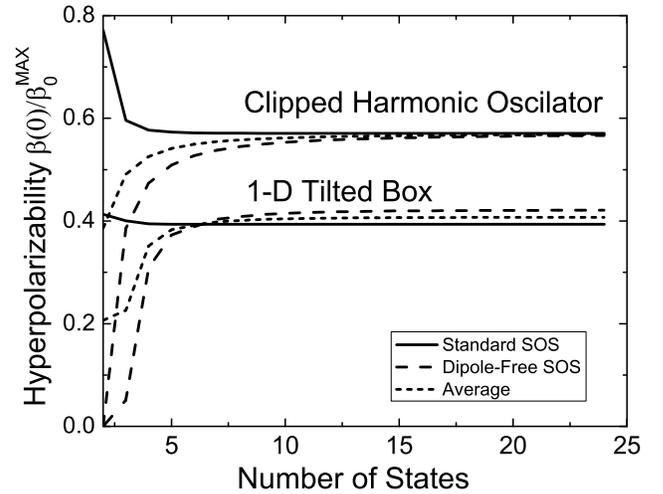}}
\caption{$\beta(0)/\beta_0^{MAX}$, the zero-frequency (off-resonance) limit of $\beta$ - normalized to the off-resonant fundamental limit - as a function of the number of excited states included in a clipped harmonic oscillator and 1D tilted Box for the standard SOS model and the dipole-free SOS expression.} \label{fig:CHOconverge}
\end{figure}
Figure \ref{fig:CHOconverge} shows the calculated off-resonant hyperpolarizability normalized to the maximum off-resonant hyperpolarizability as a function of the number of states included in the calculation.  Both theories converge to the same result as the number of states included in the sums is large, showing that the two models are identical.  Note that the standard SOS expression converges more quickly than the dipole-free expression, which suggests that the clipped harmonic oscillator is more dipolar in nature.  Presumedly, an octupolar molecule would be better modelled with the dipole-free term, resulting in faster convergence; though, there are no simple exactly soluble octupolar potentials.  The average of the two models is also shown, suggesting that a variational principle applied to a weighted average (with the weights as parameters) may yield the exact result with only a few terms.

\begin{figure}
\scalebox{1}{\includegraphics{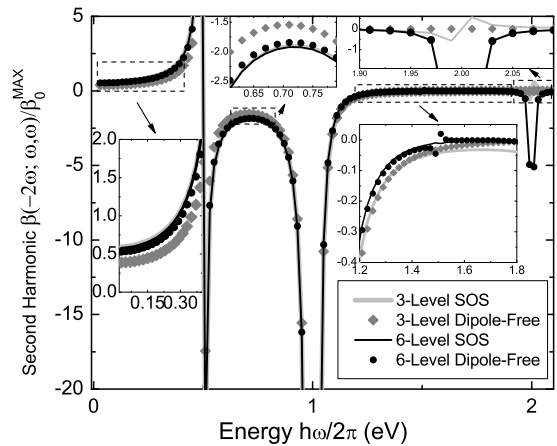}}
\caption{The normalized second harmonic hyperpolarizability ($\beta(E)/\beta_0^{MAX}$) as a function of the incident photon energy for a 3- and 6-level model of a clipped harmonic oscillator for standard and dipole-free SOS expressions.  Insets show magnified view of key regions as indicated by the dashed boxes.  The first excited state energy is arbitrarily set to $1eV$.} \label{fig:CHOdispersion}
\end{figure}
\begin{figure}
\scalebox{1}{\includegraphics{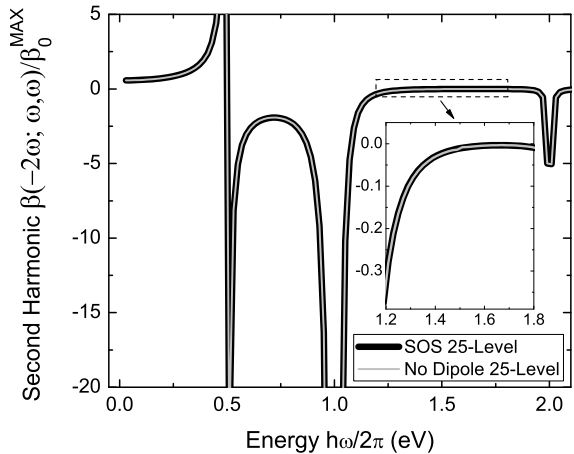}}
\caption{The normalized second harmonic hyperpolarizability ($\beta(E)/\beta_0^{MAX}$) as a function of the incident photon energy for a 25-level model of a clipped harmonic oscillator for standard and dipole-free SOS expressions.  The inset shows a magnified view.} \label{fig:CHO25dispersion}
\end{figure}

Figure \ref{fig:CHOdispersion} shows the dispersion predicted by both models for a CHO in the 3- and 6-level models for the second harmonic generation hyperpolarizability as a function of the energy of the fundamental photon.  The two theories agree well in the 6-level model except near resonance.  The insets show an expanded view of the regions in which the two theories disagree the most.  In the 25-level model(Figure \ref{fig:CHO25dispersion}), the agreement is excellent at all wavelengths,  as  expected since the CHO is an exact solution to the Schrodinger equation.

\subsection{Particle in a Tilted Box}

\begin{figure}
\scalebox{1}{\includegraphics{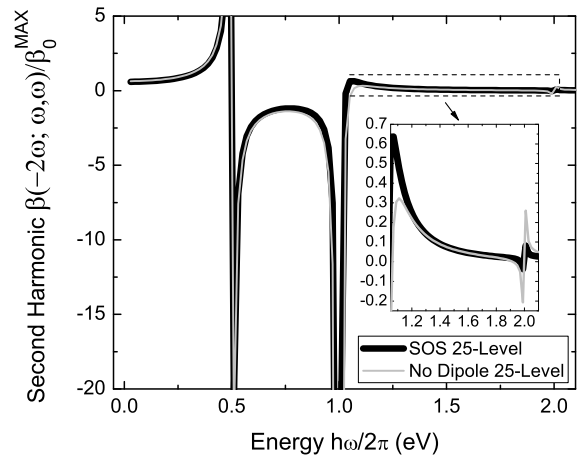}}
\caption{The normalized second harmonic hyperpolarizability ($\beta(E)/\beta_0^{MAX}$) as a function of the incident photon energy for a 25-level model of a particle in an asymmetric box for standard and dipole-free SOS expressions.  The inset shows a magnified view.} \label{fig:BOX25dispersion}
\end{figure}
Next we consider a particle in a 1-dimensional box that is perturbed by the potential $V = \epsilon x$ to make the system asymmetric.  First-order perturbation theory is used to get the wavefunction to first-order in $\epsilon$, from which the matrix elements of $x$ are calculated.  $\beta$ is calculated from these matrix elements also to first-order in $\epsilon$.  This is an interesting example because the wavefunctions, while reasonably accurate, are nevertheless only approximate.  Figure \ref{fig:CHOconverge} shows $\beta/\beta_0^{MAX}$ for the two models as a function of the number of states and Figure \ref{fig:BOX25dispersion} shows the 25-level model.  Note that the matrix elements are accurate to better than 5\%, yielding convergence of the off-resonance limit of the two 25-level models of $\beta$ to better than 7\% of each other.  However, near resonance, the two models do not agree as well quantitatively - though the qualitative features are similar.  These variations are due the inaccuracies introduced by the approximations used in calculating the wavefunctions, so it is not possible to determine which model is more accurate.  However, based on the two dispersion graphs, it is clear that the dipole free-expression and standard SOS expressions are equivalent to within the levels of uncertainty one expects from the level of approximation used.

\section{Conclusion}

In conclusion, we have derived an expression that is equivalent to the standard SOS equation for $\beta$ and $\gamma$, but does not include dipole terms.  The fact that they are identical is illustrated with the exact wavefunctions of a clipped harmonic oscillator; when the number of terms included in the sums is large, the two results converge.  However, when the {\em approximate wavefunctions} of a particle in a tilted box are used, the two expressions do not converge, illustrating how the difference can be used to estimate the uncertainty in the result.  Furthermore, such a variance may also be a sign that the wavefunctions used violate the sum rules.

The dipole-free expression is more compact; and, when truncated to a finite number of states is easier to apply to certain classes of problems, such as calculating the fundamental limits of the nonlinear susceptibility.  The dipole-free expression is a new tool for studying classes of molecules that are not well described by the truncated SOS expression.  As such, in may, for example, provide a more accurate means for analyzing the dispersion of $\beta$ for octupolar molecules.  The standard approach is to truncate the sums in Equation \ref{sumrule m<>p m,p removed final} to the first two excited states (yielding the term with numerator $x_{01} x_{12} x_{20}$).  Clearly, the dipole-free 3-level model includes more information so may be a more accurate expression for the dispersion than simply setting the dipole term in Equation \ref{beta} to zero in the standard 3-level model.

The new theory for $\beta$ and $\gamma$ presented here is therefore an additional avenue for analyzing molecules that go beyond the common dipolar push-pull paradigm, can be used to assess the accuracy of molecular orbital calculations, and sheds new light on the fundamental limits of nonlinear susceptibilities.

\section{Acknowledgements}

I thank the National Science Foundation (ECS-0354736) and Wright Paterson Air Force Base for generously supporting this work.




\clearpage

\end{document}